\documentclass{aa} 
\usepackage{natbib}
\usepackage{graphicx}
\usepackage{txfonts}
\usepackage{epsfig}
\usepackage{times}
\usepackage{rotating}
\usepackage{float}
\usepackage{setspace}
\usepackage{lscape}
\usepackage{epstopdf}
\usepackage{tabularx}
\usepackage{graphicx} 
\usepackage{txfonts}
\usepackage[caption = false]{subfig}
\usepackage[usenames]{color}
\usepackage{placeins}
\begin{document}

\title{Fossil group origins}

\subtitle{XI. The dependence of galaxy orbits on the magnitude gap}

\authorrunning{S. Zarattini et al.}

\titlerunning{The dependence of galaxy orbits on the magnitude gap}

\author{S. Zarattini\inst{1,2}, A. Biviano\inst{3,4}, J. A. L. Aguerri\inst{5,6}, M. Girardi\inst{7,3}, E. D'Onghia\inst{8}}

\institute{IRFU, CEA, Universit\'e Paris-Saclay, F-91191 Gif-sur-Yvette, France\\
	 \email{ste.zarattini@gmail.com}
\and AIM, CEA, CNRS, Universit\'e Paris-Saclay, Universit\'e Paris Diderot, Sorbonne Paris Cit\'e, F-91191 Gif-sur-Yvette, France
\and INAF-Osservatorio Astronomico di Trieste, via Tiepolo 11, I-34143 Trieste, Italy
\and IFPU Institute for Fundamental Physics of the Universe, via Beirut 2, 34014 Trieste, Italy
\and Instituto de Astrof\'isica de Canarias, calle V\'ia L\'actea s/n, E-38205 La Laguna, Tenerife, Spain
\and Departamento de Astrof\'isica, Universidad de La Laguna, Avenida Astrof\'isico Francisco S\'anchez s/n, E-38206 La Laguna, Spain
\and Dipartimento di Fisica, Universit\`{a} degli Studi di Trieste, via Tiepolo 11, I-34143 Trieste, Italy
\and Department of Astronomy, University of Wisconsin - Madison, 475 North Charter Street, Madison, WI 53706, USA}

\date{\today}

\abstract{}
  {We aim to study how the orbits of galaxies in clusters depend on the prominence of the corresponding central galaxies.}
{We divided our data set of $\sim$ 100 clusters and groups into four
  samples based on their magnitude gap between the two brightest
  members, $\Delta m_{12}$. We then stacked all the systems in each
  sample, in order to create four stacked clusters, and derive the
  mass and velocity anisotropy profiles for the four groups of
  clusters using the MAMPOSSt procedure. Once the mass profile is
  known, we also obtain the (non parametric) velocity anisotropy
  profile via the inversion of the Jeans equation.}
{In systems with the largest $\Delta m_{12}$, galaxy
  orbits are generally radial, except near the centre, where orbits
  are isotropic (or tangential when also the central galaxies are
  considered in the analysis).  In the other three samples with
  smaller $\Delta m_{12}$, galaxy orbits are isotropic or only mildly
  radial.  }
{Our study supports the results of numerical simulations that identify radial orbits of galaxies as the cause of
an increasing $\Delta m_{12}$ in groups. 
}


\authorrunning{S. Zarattini et al.}

\maketitle

\section{Introduction}

\defcitealias{Aguerri2011}{Paper~I}
\defcitealias{Mendez-Abreu2012}{Paper~II}
\defcitealias{Girardi2014}{Paper~III}
\defcitealias{Zarattini2014}{Paper~IV}
\defcitealias{Zarattini2015}{Paper~V}
\defcitealias{Kundert2015}{Paper~VI}
\defcitealias{Zarattini2016}{Paper~VII}
\defcitealias{Aguerri2018}{Paper~VIII}
\defcitealias{Corsini2018}{Paper~IX}
\defcitealias{Zarattini2019}{Paper~X}

The term ``fossil group'' (FG) was first introduced by
\citet{Ponman1994} for an apparently isolated elliptical galaxy
surrounded by an X-ray halo, with a X-ray luminosity typical of a
group of galaxies. \citet{Ponman1994} made the hypothesis that FGs
could be the fossil relics of old groups of galaxies, in which the
$L^*$ galaxies (where $L^*$ is the characteristic magnitude of the
cluster luminosity function) have had enough time to merge with the
central one (BCG). Follow-up investigations have identified companion
galaxies to the FG BCG \citep{Jones2003}, and established the
currently adopted definition of a FG. For a galaxy system to be
classified a FG, it must have an X-ray luminosity $L_X \ge 10^{42}$
$h_{50}^{-2}$ erg s$^{-1}$, and a magnitude gap, $\Delta m_{12} \geq
2$, in the $r-$band, between the BCG and the second brightest group
member within 0.5 $r_{200}$ from the BCG. With this definition, even
some clusters can enter the FG class
\citep{Cypriano2006,Zarattini2014}. 

Fossil groups are found to be transitional objects both in numerical simulations \citep{vonBenda2008,Kundert2017} and in observations \citep{Aguerri2018}.
If FGs are created by merging of $L^*$ galaxies onto the central BCG,
a mechanism is needed to enhance the central merger rate of galaxies
in FGs relative to other galaxy systems. In the standard cosmological
model, groups and clusters of galaxies form hierarchically by merging
of dark matter (DM) halos. The survival time of a subhalo accreted by
a larger one, depends on its orbit. The merger timescale
with the central halo is shorter for $L^*$ galaxies on radial
orbits than for galaxies on tangential orbits \citep[see eq. 4.2
  of][]{Lacey1993}. Subhalos on more radial orbits are
more easily destroyed, and the disrupted material is accreted onto the
central halo \citep[e.g.][]{Wetzel2011,CYK18}. Using TreeSPH simulations,
\citet{Sommer-Larsen2005} were the first to point out that galaxies in
FGs are located on more radial orbits than those in non-fossil
systems. A different orbital distribution of galaxies in fossil and
non-fossil systems could then naturally explain the increased growth
of the central galaxy in FGs at the expense of disrupted satellites
approaching on radial orbits. Moreover, \citet{D'Onghia2005} claimed that the infall
of L$^*$ galaxies along filaments with small impact parameters is required to 
explain the existence of FGs in numerical simulations.
Testing this scenario requires determining the orbits of FG galaxies.

Orbits of galaxies in non-fossil systems have been determined
observationally through the use of the Jeans equation \citep{Binney1987}
that relates the mass profile of an observed spherically-symmetric
system, $M(r)$, to the radial component of the velocity dispersion
profile, $\sigma_r(r)$, the number density profile of the tracer,
$\nu(r)$, and the velocity anisotropy profile,
\begin{equation}
\beta(r) = 1 - \frac{\sigma_\theta^2 + \sigma_\phi^2}{2\sigma_r^2},
\end{equation}
where $\sigma_\theta$, and $\sigma_\phi$, are the two tangential
components of the velocity dispersion, assumed to be identical. The
velocity anisotropy profile describes the relative content in kinetic
energy of galaxy orbits along the tangential and radial
components. For purely radial (resp. tangential) orbits $\beta=1$
(resp. $\beta=-\infty$), while $\beta=0$ correspond to isotropic
orbits. In lieu of $\beta$, a widely used parameter to describe the velocity anisotropy is $\sigma_r/\sigma_{\theta} \equiv (1-\beta)^{-1/2}$ \citep[e.g.][]{Biviano2004,Biviano2009}. For purely radial (resp. tangential) orbits $\sigma_r/\sigma_{\theta}=+\infty$ (resp. $\sigma_r/\sigma_{\theta}=0$), while $\sigma_r/\sigma_{\theta}=1$ corresponds to isotropic orbits.

Several studies found passive/red/early-type galaxies in
low-redshift clusters to follow nearly isotropic orbits, whereas star-forming/blue/late-type galaxies follow more
radially elongated orbits
\citep{Mahdavi1999,Biviano2004,Hwang2008,Munari2014,Mamon2019}.
However, this trend is not universal, since \citet{Aguerri2017} found
that early-type galaxies have more radially elongated orbits than
late-type galaxies in Abell~85. At intermediate redshifts, up to $z
\sim 1$, all cluster galaxies follow a trend of increasingly radial
orbits with increasing distance from the cluster centre
\citep{Biviano2009,Biviano2013,Biviano+16,Capasso+19}, independent of
their colour/spectral type.

Previous determinations of $\beta(r)$ have been obtained for clusters of galaxies with a sufficiently
rich spectroscopic data set, either individually \citep[e.g.][]{Biviano2013} or as stacks of several clusters \citep[e.g.][]{Biviano2009}. It is interesting to determine $\beta(r)$ for fossil systems, to learn more about their formation process, that numerical simulations suggest to be related to the orbital shape of their galaxies
\citep{Sommer-Larsen2005}. Unfortunately,
a suitable data set for fossil systems that would allow a precise determination of their $\beta(r)$ does not exist at present. 
We therefore selected a data set of 97 clusters and groups for which we measured the magnitude gap between the two brightest members, $\Delta m_{12}$, independently of whether these systems are classified as fossil or not. By stacking these systems in four bins of $\Delta m_{12}$ we can study the dependence of the orbits of their galaxies on $\Delta m_{12}$.
This is the aim of this work.  

A substantial part of the data set we use in this paper comes from the ``Fossil Group Origins'' (FOGO) project, presented in \citet{Aguerri2011}. The detailed 
study of the sample was presented in \citet{Zarattini2014} and, within the same project, we also published a study of on the properties of central galaxies in FGs \citep{Mendez-Abreu2012}, their X-ray versus optical properties \citep{Girardi2014}, the dependence on the magnitude gap of the luminosity functions \citep[LFs,][]{Zarattini2015} and substructures \citep{Zarattini2016}. The X-ray scaling relations of FGs were presented in \citet{Kundert2017}, the stellar population in FG central galaxies are analysed in \citet{Corsini2018}, and the velocity segregation of galaxies is studied in \citet{Zarattini2019}.

The structure of this paper is the following.  We describe the samples in Sect. \ref{sample}, and the methods used in our analysis in Sect. \ref{methods}. We present our results in Sect. \ref{results}, and provide our conclusions in Sect. \ref{conclusions}.

Throughout this paper, as in the rest of the FOGO papers, we adopt the following cosmology,
$H_0 = 70$ km s$^{-1}$ Mpc$^{-1}$, $\Omega_\Lambda=0.7$, and
$\Omega_{\rm M}=0.3$.

\section{Samples}
\label{sample}
In this paper we use the same data set already used in \citet{Zarattini2015} and
\citet{Zarattini2019}, that comes from the merging of two different
data sets. The first data set (S1 hereafter) comprises 34 FG
candidates proposed by \citet{Santos2007} and already analysed by the
FOGO team \citep{Aguerri2011,Zarattini2014}. The spectroscopy of S1 is
$\geq 70$\% (resp. $\geq 50$\%) complete down to $m_r = 17$
(resp. $m_r = 18$). We removed 12 systems with less than
10 spectroscopic members each. We also removed another system because
its membership assignment 
is uncertain \citep[FGS15; see][]{Zarattini2014}.
We were left with 21 systems with $z < 0.25$ and with a total of 1065 spectroscopic members. We refer the reader to \citet{Zarattini2014} for more details on S1 and
the membership selection. 

For each of the 21 S1 systems we computed $\Delta m_{12}$ \citep[see Table~1 in][]{Zarattini2014}. 
Since S1 only contains FG candidates, systems in S1 have
a high mean $\Delta m_{12} \simeq 1.5$, with only four systems with
$\Delta m_{12} < 0.5$. To determine whether the orbits of galaxies depend on
their system $\Delta m_{12}$, we need to consider another data set (that we call S2)
that includes 
systems spanning a wider range of $\Delta m_{12}$. We used the data set of 
\citet{Aguerri2007} that contain all the 88 $z < 0.1$
clusters in the catalogues of \citet{Zwicky1961}, \citet{Abell1989},
\citet{Voges1999}, and \citet{Bohringer2000}, that are also observed
in the Sloan Digital Sky Survey Data Release 4
\citep[SDSS-DR4,][]{Adelman-McCarthy2006}. 
The spectroscopic completeness of the S2 sample is
$\geq 85$\% (resp. $\geq 60$\%) down to $m_r=17$ (resp. $m_r=18$). Of the 88 available
clusters, we selected only those 76 with spectroscopically confirmed
$\Delta m_{12}$. The
total number of spectroscopic members in the S2 data set is 4338.

\begin{table*}  
\caption{Global properties of the four stacks.}
\label{tab:stack_prop}
\begin{center}
\begin{tabular}{ccrccccc}
\hline 
\noalign{\smallskip}   
\multicolumn{1}{c}{$\Delta m_{12}$} &  \multicolumn{1}{c}{$N_{{\rm cls}}$} & \multicolumn{1}{c}{$N_m$} &   
\multicolumn{1}{c}{$f(<r_{200})$} & \multicolumn{1}{c}{<$z$>} &   
\multicolumn{1}{c}{<$r_{200}$>} & \multicolumn{1}{c}{<v$_{200}$>} & \multicolumn{1}{c}{<$r_{\nu}$>}  \\  
\multicolumn{1}{c}{} &  \multicolumn{1}{c}{} & \multicolumn{1}{c}{} &   
\multicolumn{1}{c}{} & \multicolumn{1}{c}{} &   
\multicolumn{1}{c}{[Mpc]}  & \multicolumn{1}{c}{[km s$^{-1}$]} & \multicolumn{1}{c}{[$r_{200}$]}  \\  
\hline
$\le 0.5$ & 31 & 1793 & 0.8 & $0.068 \pm 0.004$ & $1.4 \pm 0.1$ & $1014 \pm  65$ & $0.53 \pm 0.05$ \\
0.5--1.0  & 23 & 1402 & 0.8 & $0.073 \pm 0.004$ & $1.3 \pm 0.1$ &  $942 \pm  55$ & $0.52 \pm 0.07$ \\
1.0--1.5  & 23 & 1416 & 0.7 & $0.084 \pm 0.012$ & $1.5 \pm 0.1$ & $1096 \pm 101$ & $0.64 \pm 0.08$ \\
$>1.5$    & 20 &  792 & 0.7 & $0.125 \pm 0.017$ & $1.2 \pm 0.1$ &  $911 \pm  81$ & $0.39 \pm 0.07$ \\
\hline
\end{tabular}
\end{center}   
\tablefoot{Column (1): sample identification.
  Column (2): number of clusters in each
    sample. Column (3): number of member galaxies. Column (4):
    fraction of members within $r_{200}$. Columns (5, 6, 7): weighted
    averages of cluster redshift, $r_{200}$, and v$_{200}$ resp. (using the
    number of member galaxies in each cluster as a weight) and their 1
    $\sigma$ uncertainties. Column (8): best-fit scale radius of the
    galaxies number density profile (in units of $r_{200}$) and its 1
    $\sigma$ uncertainty.  }
\end{table*}    


As the goal of this work is to study the dependence of $\beta(r)$ on
$\Delta m_{12}$, we divided our 97 S1+S2 galaxy systems into 4
samples in bins of $\Delta m_{12}$, chosen to ensure at least 20
systems in each bin.  The four samples contain 31 systems with
$\Delta m_{12} \le 0.5$, 23 with $0.5 < \Delta m_{12} \le 1.0$, 23
with $1.0 < \Delta m_{12} \le 1.5$, and 20 with $\Delta m_{12} > 1.5$
(see Table~\ref{tab:stack_prop}). The properties of the systems in the four samples are given in Tables~\ref{tab:sample1}, \ref{tab:sample2}, \ref{tab:sample3}, and \ref{tab:sample4}.

\section{Methodology}
\label{methods}
\begin{figure}
\includegraphics[width=0.5\textwidth]{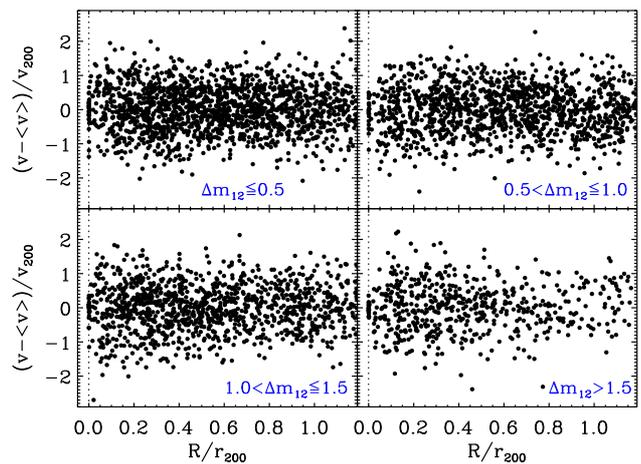}
\caption{Projected phase-space distribution of galaxies in the four samples. Top-left panel: systems with $\Delta m_{12} \le 0.5$. Top-right panel: systems with $0.5 < \Delta m_{12} \le 1.0$. Bottom-left panel: systems with $1.0 < \Delta m_{12} \le 1.5$. Bottom-right panel: systems with $\Delta m_{12} > 1.5$.}
\label{fig:caustics}
\end{figure}

\subsection{Stacking the clusters}
\label{stack}

The number of spectroscopic members in any of the clusters is too
small for allowing a robust individual cluster determination of
$\beta(r)$, with the exception of Abell~85, already analysed
by \citet{Aguerri2017}. To improve statistics,
we built stacks of clusters in each $\Delta m_{12}$
sample. In our stacking procedure we follow several previous
dynamical studies
\citep[e.g.][]{VanderMarel2000, Rines2003, Katgert2004, Biviano2016}.
The procedure relies on 
the assumption that different clusters have similar
mass profiles, differing only for the
normalisation. Such an assumption is justified by the existence of
a universal mass profile for cosmological halos \citep{Navarro1997}
and on the fact that the concentration of halo mass profiles is very
mildly mass-dependent \citep[e.g.][]{Biviano+17a}.

Following \citet{Munari2013}, we compute the virial radius\footnote{We
  define virial radius $r_{200}$ the radius of a sphere with mass
  overdensity 200 times the critical density at the cluster redshift.}
$r_{200}= \sigma_{\rm v}/(6.67 \, H_z)$, where $\sigma_{\rm v}$ is the
line-of-sight rest-frame velocity dispersion and $H_z$ is the Hubble
constant at redshift $z$. Only for one system, FGS28, we estimate its virial radius from its X-ray luminosity, $L_X$, since it does not contain enough members in its central region for a reliable $\sigma_v$ estimate (see note in Table~\ref{tab:sample4} for details).
We also compute the virial velocity
v$_{200}= 10\,H_z\, r_{200}$. We stack the individual clusters by
scaling the cluster-centric galaxy distances $R$ to the virial radius,
$R/r_{200}$, and the line-of-sight, rest-frame velocities,
$\rm{v}_{{\rm rf}} \equiv c \, (z-{\rm<}z{\rm >})/(1+{\rm <} z {\rm >})$, where $c$ is the speed of light and
<$z$> is the cluster mean redshift, to the virial velocity,
$\rm{v}_{rf}/\rm{v}_{200}$.

The properties of the four stacks are computed as the weighted
averages of the properties of the clusters in each sample, using
the number of member galaxies as weights, and are presented in Table
\ref{tab:stack_prop}. The four stacks have very similar mean
redshifts, while the mean $r_{200}$ values are marginally different
among each others.  The projected phase-space distributions of galaxies
in the four samples is shown in Fig. \ref{fig:caustics}.

\subsection{MAMPOSSt}
\label{mass_profile}
When the mass profile of a cluster is derived from the kinematics of
its galaxies (as in this work) using the Jeans equation
\citep[e.g.][]{vanderMarel94}, the solutions for $M(r)$ and $\beta(r)$
are degenerate with respect to the usually adopted osbervables, the
projected number-density and velocity-dispersion profiles
\citep[e.g.][]{Walker+09}. The MAMPOSSt technique of \citet{Mamon2013}
has been shown to partially break this degeneracy. It estimates $M(r)$ and
$\beta(r)$ in a parametrised form, by performing a maximum likelihood
fit to the full distribution of galaxies in the projected phase
space.

In our analysis we consider three models for $M(r)$:
\begin{itemize}
\item the Navarro, Frenk, and White profile \citep[NFW,][]{Navarro1997},
\begin{equation}
M_{NFW}(r) = M_{200}\frac{{\rm ln}(1+r/r_{-2})-r/r_{-2}(1+r/r_{-2})^{-1}}{{\rm ln}(1+c_{200})-c_{200}/(1+c_{200})},
\end{equation}
where $M_{200} \equiv 100 \, H_z^2 \, r_{200}^3/G$, and $H_z$ is the
Hubble constant at the system redshift, $c=r_{200}/r_{-2}$ is the
concentration of $M(r)$, and $r_{-2}$ is the scale radius, defined as
the radius where the NFW profile has a logarithmic slope of -2
\citep{Navarro2004}.
\item The Einasto profile \citep{Einasto1965,Navarro2004},
\begin{equation}
M_E(r) = M_{200} \frac{P[3m,2m(r/r_{-2})^{1/m}]}{P[3m,2m(r_{200}/r_{-2})^{1/m}]},
\end{equation}
where $P(a,x)$ represents the regularised incomplete gamma function, and where we fix $m=5$, which represents well cluster-size halos in numerical simulations \citep{MBM10}.
\item The Burkert profile \citep{Burkert1995,Biviano2013},
\begin{equation}
\begin{split}
M_B(r) = M_{200} \, \{\ln[1+(r/r_B)^2]+2 \ln(1+(r/r_B) \, -\\
-\, 2 \arctan(r/r_B)\}  \times \,\{\ln[1+(r_{200}/r_B)^2]\, + \\
+\, 2 \ln(1+(r_{200}/r_B)-2 \arctan(r_{200}/r_B)\}^{-1},
\end{split}
\end{equation}
\end{itemize}
where $r_B$ is the scale radius of the model.
All these models have two free parameters, $r_{-2}$ and
$r_{200}$. However, the four stacks on which we run MAMPOSSt, have the
observables already defined in virial units, $R/r_{200}$ and
$\rm{v}_{rf}/\rm{v}_{200}$ (see Sect.~\ref{stack}), so $r_{200}$ is no
longer a free parameter.  In Sect.~\ref{results} we show that if we
allow $r_{200}$ as a free parameter in the MAMPOSSt analysis, the
best-fit values are consistent with the mean values reported in
Table~\ref{tab:stack_prop}, and the likelihood of the MAMPOSSt
best-fit does not improve significantly  with respect to keeping
$r_{200}$ fixed.

We consider five different models for $\beta(r)$: 
\begin{itemize}
\item ``C'': constant anisotropy with radius, $\beta = \beta_c$.
\item ``T'': from \citet{Tiret2007},
\begin{equation}
\beta_T(r)=\beta_{\infty}\frac{r}{r+r_{-2}},
\end{equation}
where $\beta_{\infty}$ is the anisotropy value at large radii.
\item ``O'': from \citet{Biviano2013},
\begin{equation}
\beta_O(r)=\beta_{\infty}\frac{r-r_{-2}}{r+r_{-2}}.
\end{equation}
\item ``ML'': from \citet{Mamon2005},
\begin{equation}
\beta_{ML}(r)=\frac{1}{2}\, \frac{r}{r+r_\beta},
\end{equation}
where $r_\beta$ is the anisotropy radius.
\item ``OM'': from \citet{Osipkov79} and \citet{Merritt1985}, 
\begin{equation}
\beta_{OM}(r)=\frac{r^2}{r^2+r_{\beta}^2}.
\end{equation}
\end{itemize}
All these $\beta(r)$ models have one free
parameter each ($\beta_C, \beta_{\infty},$ or $r_{\beta}$).

We run MAMPOSSt in the so-called Split mode \citep{Mamon2013}, that is
we use an external maximum-likelihood analysis to determine the value
of the scale radius of the galaxies number density profile,
$r_{\nu}$. We fit the radial distributions of the galaxies in each
stack with NFW models (in projection), taking into account the
correction for sample incompleteness as in \citet{Zarattini2019}.  The
best fit values for $r_{\nu}$ are given in
Table~\ref{tab:stack_prop}. The $\Delta m_{12} > 1.5$ sample has a slightly more concentrated distribution
of galaxies than the other three samples.

\subsection{Inversion of the Jeans equation}
\label{inversion}
While MAMPOSSt is able to constrain $M(r)$ and $\beta(r)$, the
constraints are specific to the set of models that are considered (see
the previous Sect.). There is a vast literature on the modelisation
of cluster $M(r)$ \citep[e.g.][and references
  therein]{Ludlow+13,Pratt+19}. On the other hand, less is known from
numerical simulations and observations about the shape of $\beta(r)$
in galaxy systems, and a large variance among different systems has
been suggested \citep[see Fig.~1 in][]{Mamon2013}. Our choice of models
for MAMPOSSt could therefore be adequate to describe $M(r)$, but not
perhaps to describe $\beta(r)$. To confirm that our $\beta(r)$
modelisation is not too restrictive we use the $M(r)$ determined by
the MAMPOSSt analysis, to directly invert the Jeans equation and
derive $\beta(r)$ in an (almost) non-parametric way. For this, we
follow the method of \citet{Binney1987} in the implementations of
\citet{Solanes1990} and \citet{DM92}.

Our procedure is the following. We fix $M(r)$ to the MAMPOSSt
solution.  The two observables we need to consider are the number
density and velocity dispersion profiles.  We apply the LOWESS
technique \citep[see, e.g.,][]{Gebhardt+94} to smooth these
profiles. The number density profile is then de-projected numerically,
\citep[using Abel's equation, see][]{Binney1987}. Since the equations
to be solved contain integrals up to infinity, we extrapolate the
profiles to a large-enough radius -- we find 30 Mpc to be sufficient
for our results to be stable. The extrapolations are performed as in
\citet{Biviano2013}.

Uncertainties in the $\beta(r)$ profiles are estimated by performing
the Jeans inversion on 100 bootstrap resamplings of the original
datasets.


\section{Results}
\label{results}
\subsection{MAMPOSSt}
We apply MAMPOSSt to the four samples of Table~\ref{tab:stack_prop},
limiting each data set to the region $0.05 \, \rm{Mpc} \leq R \leq
r_{200}$, since the inner region, $R < 0.05$ Mpc, is dominated by the
BCG, where our parametrisation of $M(r)$ may not work because the
total mass is no longer DM-dominated \citep[e.g.][]{BS06}, while the
outer region, $R > r_{200}$, may not have reached dynamical equilibrium
yet.

\begin{table*}  
\caption{Results.} 
\label{tab:results}
\begin{center}
\begin{tabular}{cccccc}
\hline 
\noalign{\smallskip}   
\multicolumn{1}{c}{$\Delta m_{12}$} & \multicolumn{1}{c}{$M(r)$} &
\multicolumn{1}{c}{$\beta(r)$} & \multicolumn{1}{c}{$r_{-2}$} &
\multicolumn{1}{c}{$\beta(r_{200}/4)$} &  \multicolumn{1}{c}{$\beta(r_{200})$}  \\  
\multicolumn{1}{c}{} &  \multicolumn{1}{c}{model} &
\multicolumn{1}{c}{model} &   \multicolumn{1}{c}{[Mpc]} &
\multicolumn{1}{c}{} & \multicolumn{1}{c}{}  \\  
\multicolumn{1}{c}{(1)} &  \multicolumn{1}{c}{(2)} &
\multicolumn{1}{c}{(3)} &   
\multicolumn{1}{c}{(4)} & \multicolumn{1}{c}{(5)} &   
\multicolumn{1}{c}{(6)}  \\  
\hline
\multicolumn{6}{c}{MAMPOSSt: minimum BIC} \\
\hline
\noalign{\smallskip}   
$\le 0.5$ & Einasto & C & $0.8 \pm 0.2$ & $1.2 \pm 0.1$ & $1.2 \pm 0.1$ \\
0.5--1.0  & Einasto & C & $1.0 \pm 0.3$ & $1.3 \pm 0.1$ & $1.3 \pm 0.1$ \\
1.0--1.5  & Einasto & C & $0.7 \pm 0.2$ & $1.3 \pm 0.1$ & $1.3 \pm 0.1$ \\
$> 1.5$   & Burkert & T & $0.3 \pm 0.1$ & $1.4 \pm 0.1$ & $2.1 \pm 0.4$ \\
\hline
\multicolumn{6}{c}{MAMPOSSt: minimum BIC for NFW models} \\
\hline
$\le 0.5$ & NFW & C & $0.7 \pm 0.2$ & $1.2 \pm 0.1$ & $1.2 \pm 0.1$ \\
0.5--1.0  & NFW & C & $0.9 \pm 0.2$ & $1.3 \pm 0.1$ & $1.3 \pm 0.1$ \\
1.0--1.5  & NFW & C & $0.6 \pm 0.2$ & $1.3 \pm 0.1$ & $1.3 \pm 0.1$   \\
$> 1.5$   & NFW & T & $0.5 \pm 0.1$ & $1.3 \pm 0.1$ & $1.7 \pm 0.3$   \\
\hline
\multicolumn{6}{c}{MAMPOSSt: $\hat{L}$-weighted average of all models} \\
\hline
$\le 0.5$ & all & all & $0.6 \pm 0.2$ & $1.1 \pm 0.1$ & $1.1 \pm 0.1$ \\
0.5--1.0  & all & all & $0.7 \pm 0.3$ & $1.1 \pm 0.1$ & $1.2 \pm 0.1$ \\
1.0--1.5  & all & all & $0.6 \pm 0.3$ & $1.2 \pm 0.1$ & $1.2 \pm 0.1$  \\
$> 1.5$   & all & all & $0.4 \pm 0.2$ & $1.3 \pm 0.3$ & $1.5 \pm 0.3$ \\
\hline
\multicolumn{6}{c}{Jeans equation inversion} \\
\hline
\noalign{\smallskip}   
$\le 0.5$ & Einasto & none & 0.8 & $1.1 \pm 0.1$ & $1.3 \pm 0.1$ \\
0.5--1.0  & Einasto & none & 1.0 & $1.5 \pm 0.1$ & $1.4 \pm 0.1$ \\
1.0--1.5  & Einasto & none & 0.7 & $1.4 \pm 0.2$ & $1.6 \pm 0.1$ \\
$> 1.5$   & Burkert & none & 0.3 & $1.4 \pm 0.2$ & $1.9 \pm 0.4$ \\
\hline
\end{tabular}
\end{center}   
\tablefoot{Column (1): sample identification. Column (2): $M(r)$
  model. Column (3): $\beta(r)$ model. Column (4): $r_{-2}$ and and 1
  $\sigma$ uncertainty; Column (5, 6): values of $\beta(r)$ calculated
  at two radii, $r_{200}/4$ and $r_{200}$, and their 1 $\sigma$
  uncertainties.  Results are presented for the combination of $M(r)$
  and $\beta(r)$ models that return the minimum BIC value, for the NFW
  $M(r)$ models that return the minimum BIC values, and for the
  weighted average of all considered models, where the weights are
  proportional to the MAMPOSSt likelihoods $\hat{L}$. The quoted
  uncertainties are marginalised errors obtained from a MCMC analysis
  in the first two sets of results, and from the variance among the
  different model results in the third set of results. The results for
  the Jeans inversion are obtained by fixing $r_{-2}$, thus no error
  is reported for this quantity. }
\end{table*}    

We compare the MAMPOSSt solutions obtained from the 15 combinations of
the three $M(r)$ and the five $\beta(r)$ models (see
Sect.~\ref{mass_profile}) using the {\it Bayesian Information
  Criterion} \citep[BIC][]{Schwarz78},
\begin{equation}
{\rm BIC} = N_{\rm{pars}} \ln N_{\rm{data}} \,  - 2 \ln (\hat{L}),
\end{equation}
where $N_{\rm{data}}$ is the sample size, and $N_{\rm{pars}}$ is the
number of free parameters used in the model and $\hat{L}$ is the MAMPOSSt-derived likelihood. The BIC values obtained
using two free parameters ($r_{-2},$ and the $\beta(r)$ parameter) are
systematically lower than the BIC values obtained using three free parameters
(i.e. adding $r_{200}$ as a free parameter) for all combinations of
$M(r)$ and $\beta(r)$ models. This means there is no
statistical advantage of adding $r_{200}$ as a free parameter in our
analysis, presumably because the stack sample observables are already
in normalised units with respect to $r_{200}$ and
$\rm{v}_{200}$. Anyway, we checked that the best-fit $r_{200}$ values
obtained by MAMPOSSt in the 3-free parameters runs, are consistent
with the weighted mean values of $r_{200}$ resulting from the cluster
stacking procedure (listed in Table~\ref{tab:stack_prop}; see also
Sect.~\ref{stack}).

\begin{figure}
\includegraphics[width=0.5\textwidth]{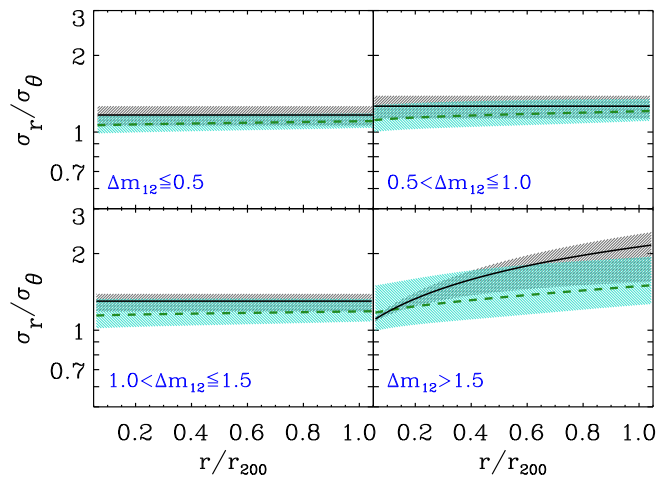}
\caption{MAMPOSSt estimates of the velocity anisotropy profile, $\sigma_r/\sigma_{\theta}$, for the four samples. Black
  curve and grey shading: minimum-BIC solution and 1 $\sigma$ confidence
  region estimated by the MCMC analysis. Green dashed curve and
  turquoise shading: weighted average and dispersion of the MAMPOSSt
  results from all different combinations of $M(r)$ and $\beta(r)$
  models, using the MAMPOSSt likelihoods as weights. See also
  Table~\ref{tab:results}.  Top-left panel: systems with $\Delta
  m_{12} \le 0.5$. Top-right panel: systems with $0.5 < \Delta m_{12}
  \le 1.0$. Bottom-left panel: systems with $1.0 < \Delta m_{12} \le
  1.5$. Bottom-right panel: systems with $\Delta m_{12} > 1.5$.}
\label{fig:beta_mamp}
\end{figure}

The main results of the MAMPOSSt analysis are given in
Table~\ref{tab:results}. We list the best-fit values of $r_{-2}$ and
the values of $\beta(r)$ at two characteristic radii ($r_{200}/4$ and
$r_{200}$). These values are listed for the combination of $M(r)$ and
$\beta(r)$ models that give the minimum BIC values for each of the
four samples. The minimum-BIC solutions for the three smaller 
$\Delta m_{12}$ samples are obtained using the Einasto mass profile and
the constant anisotropy profile. On the other hand, for the $\Delta m_{12} > 1.5$ sample
the minimum-BIC solution is obtained using the Burkert mass model and 
the T anisotropy model. The listed uncertainties
are marginalised errors obtained from a Markov chain Monte Carlo (MCMC) analysis.

The $\Delta m_{12} > 1.5$ sample differs from the other three not only for the
different minimum-BIC models, but also for the larger value of
$\beta(r_{200})$, and for the smaller value of $r_{-2}$. The smaller
$r_{-2}$ implies a higher concentration of the mass distribution, as
already found for the galaxy distribution in Sect. \ref{mass_profile}. A
higher mass concentration for systems with large magnitude gaps is predicted by
cosmological numerical simulations \citep{Ragagnin+19}.  However, the
smaller $r_{-2}$ we find for  $\Delta m_{12} > 1.5$ systems probably compensates for
the fact that the Burkert mass model is cored at the centre, unlike
the Einasto model. In fact, when we force the NFW $M(r)$ model to all
the samples, the $r_{-2}$ values of the four samples are not much
different (see the second set of results in
Table~\ref{tab:stack_prop}). On the other hand, the $\beta(r_{200}$ value of the $\Delta m_{12} > 1.5$ sample is larger than the corresponding values of the
the other three samples, independently of the $M(r)$ model.

The third set of results shown in Table~\ref{tab:results} are the
weighted average MAMPOSSt results of all models combinations, using
the MAMPOSSt likelihoods $\hat{L}$ as weights. The quoted errors on
the parameters are the weighted variance. Also for this set of
results, it is confirmed that the $\Delta m_{12} > 1.5$ sample has a higher
$\beta(r_{200})$ value compared to the other
three samples, although the difference is less significant than for
the minimum-BIC, and the minimum-BIC NFW sets of results.

\begin{figure}
\includegraphics[width=0.5\textwidth]{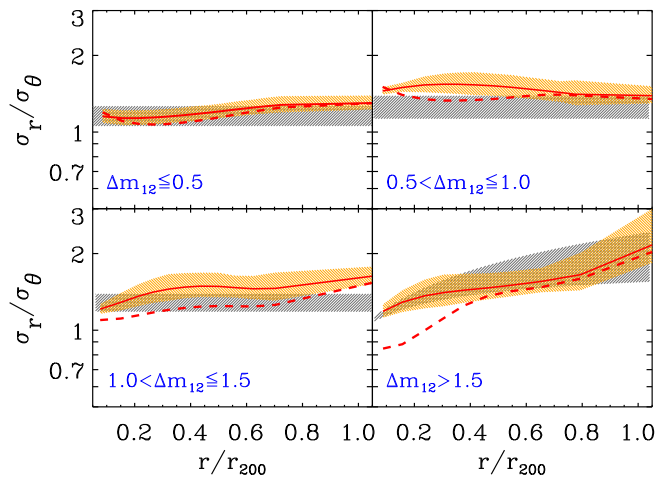}
\caption{Red solid curves and orange shadings: velocity anisotropy profile $\sigma_r/\sigma_{\theta}$ and 1
  $\sigma$ confidence regions (estimated from 100 bootstrap
  resampling) for the four samples, obtained from the Jeans equation
  inversion using the minimum-BIC MAMPOSSt $M(r)$ (see
  Table~\ref{tab:results}).  The dashed red curves indicate the
  solutions obtained including galaxies in the central $<0.05$ Mpc
  regions. For comparison, the grey shadings reproduce the 1 $\sigma$
  confidence regions of the MAMPOSSt solutions shown in
  Fig.~\ref{fig:beta_mamp}. Top-left panel: systems with $\Delta
  m_{12} \le 0.5$. Top-right panel: systems with $0.5 < \Delta m_{12}
  \le 1.0$. Bottom-left panel: systems with $1.0 < \Delta m_{12} \le
  1.5$. Bottom-right panel: systems with $\Delta m_{12} > 1.5$.}
\label{fig:beta_jeans}
\end{figure}

In Fig.~\ref{fig:beta_mamp} we display the four samples velocity anisotropy profiles, $\sigma_r/\sigma_{\theta}$,
corresponding to the first and third sets of results of
Table~\ref{tab:results}. We do not show the velocity anisotropy models
obtained by forcing the NFW $M(r)$ for the sake of clarity of the plot
-- anyway, they are quite similar to those of the minimum-BIC models.
The velocity anisotropy of the $\Delta m_{12} > 1.5$  sample is increasing with radius, indicating more
radial orbits in the outer regions than the other three samples.

The differences we found are larger than $1 \sigma$, but smaller than $3 \sigma$. There
is therefore only tentative evidence for the presence of more radial orbits in systems with large magnitude gap.  A larger data set is needed to provide a more solid statistical basis to our result, and eventually to extend it to a sample of pure fossil systems.

\subsection{Jeans equation inversion}
Given the best-fit MAMPOSSt $M(r)$ models and the observables, namely
the galaxies number-density and velocity-dispersion profiles, we now
perform the inversion of the Jeans equation to determine $\beta(r)$ in
a non-parametric form. This procedure allows us to free the
determination of $\beta(r)$ from the constraints imposed by the choice
of models used in the MAMPOSSt analysis. Also in this case we limit
the analysis to the region $0.05 \, \rm{Mpc} \leq R \leq r_{200}$.

The results of the Jeans inversion analysis are shown in
Fig.~\ref{fig:beta_jeans}. The results are similar to those obtained
with MAMPOSSt. The marginal differences (always within 20\%)
between the $\sigma_r/\sigma_{\theta}$ profiles obtained
by the two methods can be attributed to the limited number of $\beta(r)$ models
considered in MAMPOSSt.

There appears to be a trend of increasing $\beta(r)$ at large radii
with increasing $\Delta m_{12}$. This is confirmed by the values of
$\beta(r_{200})$ reported in Table~\ref{tab:results}.

Near the centre, the situation is less clear. To better investigate
the inner region, we repeat the Jeans inversion analysis by extending
the analysed region to $0.0 \leq R \leq r_{200}$. The results are
shown in Fig.~\ref{fig:beta_jeans} (dashed lines). Including the
galaxies near the centre leads to decreasing velocity anisotropy near
the centre in all samples, but the decrease is stronger in the systems
with higher $\Delta m_{12}$. A possible explanation for this behaviour
resides in the velocity segregation of BCGs that is stronger
for systems with  higher $\Delta m_{12}$, as found by
\citet{Zarattini2019}. Dynamical friction can decrease the
velocities of galaxies but also make their orbits more isotropic if not tangential.

\subsection{Systematics}
Here we examine possible systematics affecting our result.
 \begin{figure}
\includegraphics[width=0.5\textwidth]{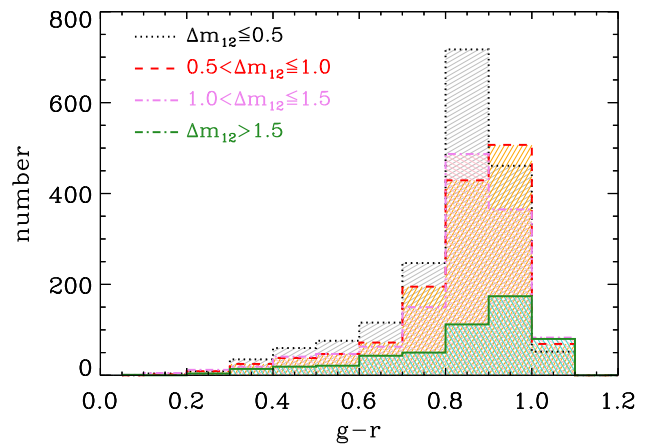}
\caption{$g-r$ colour distribution of the galaxies in the four samples. Dotted black histogram and grey shading: $\Delta m_{12} \leq 0.5$. Dashed red histogram and orange shading: $0.5 < \Delta m_{12} \leq 1.0$. Dash-dotted violet histogram and pink shading: $1.0 < \Delta m_{12} \leq 1.5$. Solid green histogram and turquoise shading:  $\Delta m_{12} > 1.5$.}
\label{fig:col_dist}
\end{figure}

\begin{figure}
\includegraphics[width=0.5\textwidth]{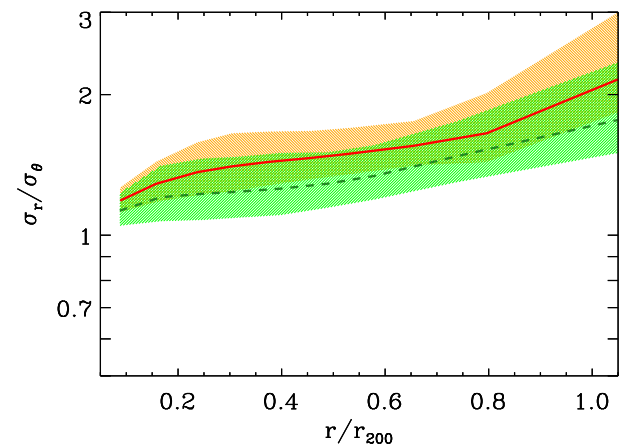}
\caption{Velocity anisotropy profile $\sigma_r/\sigma_{\theta}$ and 1
  $\sigma$ confidence regions for the systems with $\Delta m_{12} > 1.5$. Solid red line and orange shading: based on all 20 systems (same as bottom-right panel in Fig.~\ref{fig:beta_jeans}); dashed green line and green shading: based on 17 systems, excluding the three richest.}
\label{fig:beta_poor}
\end{figure}

In low-$z$ clusters, late-type/blue galaxies are observed to have more
radially anisotropic orbits than early-type/red galaxies
\citep[e.g.]{Biviano2004,Munari2013,Mamon2019}. If the $\Delta m_{12} > 1.5$ systems
contain a a larger fraction of late-type/blue galaxies compared to the
systems that compose the other three samples, this could explain the
higher values of $\beta$ at large radii.

Lacking detailed studies of galaxy populations as a function of
$\Delta m_{12}$ in the literature, we here determine the galaxies
$g-r$ colour distribution in the four stacks. These are shown in
Fig.~\ref{fig:col_dist}. It can be seen that systems in the $\Delta m_{12} > 1.5$ bin do not show a larger amount of late-type/blue galaxies. If anything, they contain more red galaxies, mostly because of the deep spectroscopic follow-up of
the S1  data set, that contributes most systems in the $\Delta m_{12} > 1.5$ bin. The S1 data set includes many dwarf galaxies, up to three magnitudes fainter than SDSS spectroscopy. Dwarf galaxies in clusters/groups are mostly early-type \citep[e.g.][]{Jerjen1997,Lisker2013}, and therefore occupy the red tail of the $g-r$ distribution.

Different $\beta(r)$ have also been reported for galaxies of different
stellar masses \citep{Annunziatella2016} or luminosity \citep{Aguerri2017}. To check if a different
magnitude distribution could be at the origin of the different
$\beta(r)$ seen for the $\Delta m_{12} > 1.5$ stack, we repeat the MAMPOSSt
analysis using only galaxies with $r \le 17.77$. This is the
magnitude limit of the SDSS spectroscopy, and effectively exclude
the tail of red dwarf galaxies from the two highest-$\Delta m_{12}$
samples, while leaving almost unchanged the other two samples. We
find that  the results for the $\beta(r)$ of the four stacks do not change significantly when applying the magnitude cut $r \le 17.77$, with $\beta(r_{200})$ changing by $<5$\%.

The number of members is very different in different systems of our data set. To check that our result is not driven by a few very rich systems, we removed from the $\Delta m_{12}>1.5$ sample the three richest clusters, FGS03, FGS27 and FGS30 that together contain almost 1/3 of all members in their sample (see Table~\ref{tab:sample4}). We performed the full analysis on the remaining sample of 17 systems. The resulting $\sigma_r/\sigma_{\theta}$ profile is very similar to the original one based on all 20 systems (see Fig.~\ref{fig:beta_poor}).

The S1 and S2 data sets cover different $z$ ranges, their median $z$ are 0.16 and 0.07, respectively. This difference reflects in an increase of the <$z$> of the four samples with increasing $\Delta m_{12}$, since the higher-$\Delta m_{12}$ systems are mostly from the S1 data set. To check for a $z$-dependence of $\beta(r)$ we divided our $\Delta m_{12}>1.5$ sample in two subsamples, of ten clusters at $z<0.12$ and another ten at $z>0.12$. After performing the dynamical analysis separately on the two subsamples, we found no significant difference in their $\beta(r)$, but the uncertainties are large due to the small size of the subsamples, so this test cannot be considered very significant. 

There is independent support against the hypothesis that the <$z$> difference across the four samples can be the reason for the observed difference in $\beta(r)$. The <$z$> range across the four samples correspond to 0.7 Gyr in cosmic time. This is only 25\% of the dynamical time for a typical system of galaxies at $z \sim 0.1$ \citep{Sarazin1986}, and it is unlikely that galaxies could modify their orbits in such a short time. Moreover, there is no observational
evidence for orbital evolution of cluster galaxies across the much larger cosmic time span
from $z=1.32$ to $z=0.26$ \citep[corresponding to 6 Gyr,][]{Capasso2019}.

\section{Discussion and conclusions}
\label{conclusions}

We analysed a data set of 97 galaxy clusters and groups to study the
dependence of $\beta(r)$, and therefore of the orbital distribution of
galaxies, on $\Delta m_{12}$.  We split our data set in four samples
of different $\Delta m_{12}$. We then stacked together the systems
in each of the four samples.  We
then run MAMPOSSt to derive the mass and the (parametric) anisotropy
profiles of the four samples. Finally, with the mass profiles
obtained from MAMPOSSt, we perform the inversion of the Jeans
equation, allowing us to determine $\beta(r)$ in a model-independent
way.

We find that $\beta(r)$ shows a steeper dependence on $r$ for the
systems with $\Delta m_{12} > 1.5$ than for the other three
samples with smaller $\Delta m_{12}$. The orbits of galaxies in the $\Delta m_{12} > 1.5$ stack are more radial (with marginal significance, at more than $1\sigma$ level) at large radii ($r \gtrsim 0.8 \, r_{200}$) than in
systems with smaller $\Delta m_{12}$. In the central regions the
orbits of galaxies are nearly isotropic in all stacks, or
even tangential at radii $<0.05$ Mpc in the  $\Delta m_{12} > 1.5$ stack. The
tangential orbits found in the very central regions of these systems are related
to the observed velocity segregation \citep{Zarattini2019} in the
same region, and can be interpreted as an effect of dynamical
friction slowing down galaxies that approach their cluster centre.

Dynamical friction is thought to be more efficient for galaxies on
radial orbits \citep{Lacey1993}. As galaxies lose their kinetic energy
due to dynamical friction, they can more easily merge with the central
galaxy, and this is the process suggested by \citet{Sommer-Larsen2005}
for the formation of large magnitude gaps in galaxy systems.

In \citet{Zarattini2015} we studied the dependence of the luminosity function (LF) on the magnitude gap. We found that systems with  $\Delta m_{12} > 1.5$ are missing not only L$^*$, but also dwarf galaxies. In fact, the faint end of their LFs is clearly flatter than that of  $\Delta m_{12} < 1.5$ systems. Moreover, \citet{Adami2009} found that dwarf galaxies  in Coma are located in radial orbits even in the central region of the cluster. Thus, we suggest that also the lack of dwarf galaxies in FGs could be linked to radial orbits. Unfortunately, deeper data are required to study the orbits of dwarf galaxies in FGs, studies that could be done in the near future with new wide-field spectroscopic facilities (e.g. the WEAVE spectrograph).

Galaxy systems are thought to evolve from an initial phase of rapid
collapse characterised by isotropisation of galaxy orbits, to a phase
of slow accretion characterised by radial orbits \citep{Lapi2011}.
Major mergers operate in the same way as the initial phase of rapid
collapse, introducing dynamical entropy in the system, and
transferring angular momentum from clusters colliding off-axis to
galaxies, leading to orbit isotropisation.  The fact that $\Delta m_{12}>1.5$ systems
have more galaxies on radial orbits than smaller-$\Delta m_{12}$ systems therefore
suggests a difference in the time since last major merger,
and this supports the conclusions of \citet{Kundert2017} based on
cosmological simulations.

However, although \citet{D'Onghia2005} found that radial orbits are required for the formation of FGs (see their Eq. 3), there is no clear evidence of correlation between the type of orbits and the magnitude gap in numerical simulations. Our observational confirmation of the presence of radial orbits in systems with large magnitude gaps could be a boost for the analysis of future simulations in order to explain such a difference.

In summary, our study is a first observational confirmation that
galaxies in systems with large  $\Delta m_{12}$ have more radially elongated orbits than
galaxies in systems with small $\Delta m_{12}$. 
Our samples contain a mix of pure FGs and normal systems. A substantial increase of the spectroscopic data set for FGs is needed to check if our results also apply for these systems as a separated class.

\begin{acknowledgements}
We thank the anonymous referee for his/her useful comments. S.Z. acknowledges funding from the European Research Council under the European Union's Seventh
Framework Programme (FP7/2007-2013) / ERC grant agreement n. 340519.
M.G. acknowledges the support from the grant MIUR PRIN 2015
``Cosmology and Fundamental Physics: illuminating the Dark
Universe with Euclid''. A.B. acknowledges the hospitality of the Instituto de Astrof\'isica de Canarias during a workshop where the foundations of this project were set.
\end{acknowledgements}

\bibliography{bibliografia_ABi}

\begin{appendix}
\section{Sample properties}
Here we present the main properties for the four samples of systems with different $\Delta m_{12}$. Some of them were already published in \citet{Zarattini2014} and \citet{Aguerri2007}, respectively.

\begin{table*}
\caption{Global properties of the $\Delta m_{12} \leq 0.5$ sample.}
\label{tab:sample1}
\centering
\begin{tabular}{lrrrrrr}
\hline 
\noalign{\smallskip}   
\multicolumn{1}{l}{Name} & \multicolumn{1}{c}{R.A.} &
\multicolumn{1}{c}{Dec} & \multicolumn{1}{c}{$N_m$} & \multicolumn{1}{c}{z} &  
\multicolumn{1}{c}{$r_{200}$} & \multicolumn{1}{c}{$\Delta m_{12}$} \\  
& & & & & \multicolumn{1}{c}{[Mpc]} &  \\  
\hline
         ABELL2092 & 233.333000 &  31.212000 &  45 & 0.067 & 0.95 & 0.00 \\
   ZwCl1316.4-0044 & 199.807000 &  -0.987810 &  39 & 0.083 & 1.14 & 0.01 \\
         ABELL1452 & 180.780000 &  51.675200 &  25 & 0.062 & 1.10 & 0.06 \\
         ABELL1142 & 165.239000 &  10.505500 &  61 & 0.035 & 1.17 & 0.07 \\
         ABELL1270 & 172.175000 &  54.172300 &  52 & 0.069 & 1.17 & 0.09 \\
    RXJ1053.7+5450 & 163.402000 &  54.868000 &  61 & 0.072 & 1.37 & 0.09 \\
   ZwCl1730.4+5829 & 261.862000 &  58.516500 &  47 & 0.028 & 1.03 & 0.12 \\
         ABELL2169 & 243.550000 &  49.120300 &  49 & 0.058 & 1.08 & 0.14 \\
             FGS06 & 131.235964 &  42.976592 &  28 & 0.054 & 0.75 & 0.20 \\
         ABELL1003 & 156.257000 &  47.841900 &  31 & 0.063 & 1.28 & 0.22 \\
         ABELL1066 & 159.778000 &   5.209770 &  98 & 0.069 & 1.71 & 0.23 \\
         ABELL1205 & 168.334000 &   2.546670 &  83 & 0.076 & 1.83 & 0.25 \\
            WBL238 & 146.689000 &  54.426900 &  67 & 0.047 & 1.26 & 0.27 \\
         ABELL0757 & 138.282000 &  47.708400 &  40 & 0.051 & 0.85 & 0.33 \\
         ABELL2149 & 240.367000 &  53.947400 &  34 & 0.065 & 0.95 & 0.33 \\
         ABELL0602 & 118.361000 &  29.359500 &  83 & 0.061 & 1.73 & 0.35 \\
         ABELL2018 & 225.283000 &  47.276600 &  44 & 0.087 & 1.30 & 0.35 \\
  MACSJ0810.3+4216 & 122.597000 &  42.273900 &  41 & 0.064 & 1.05 & 0.35 \\
   RXCJ0137.2-0912 &  24.314100 &  -9.197610 &  46 & 0.041 & 0.95 & 0.35 \\
   RXCJ1424.8+0240 & 216.198000 &   2.664440 &  20 & 0.054 & 1.12 & 0.37 \\
             FGS09 & 160.760712 &   0.905070 &  67 & 0.125 & 1.73 & 0.40 \\
         ABELL1436 & 179.860000 &  56.403700 &  86 & 0.065 & 1.47 & 0.41 \\
   ZwCl1215.1+0400 & 184.421000 &   3.655840 & 129 & 0.077 & 1.96 & 0.41 \\
         ABELL1616 & 191.851000 &  54.987000 &  42 & 0.083 & 1.16 & 0.42 \\
         ABELL1552 & 187.548000 &  11.743800 &  17 & 0.088 & 0.90 & 0.44 \\
         ABELL1507 & 183.703000 &  59.906200 &  42 & 0.060 & 0.79 & 0.45 \\
         ABELL1016 & 156.782000 &  11.010700 &  27 & 0.032 & 0.54 & 0.46 \\
         ABELL1885 & 213.431000 &  43.644800 &  23 & 0.089 & 1.11 & 0.47 \\
         ABELL2255 & 258.120000 &  64.060800 & 267 & 0.080 & 1.81 & 0.47 \\
   RXCJ1121.7+0249 & 170.386000 &   2.887250 &  68 & 0.049 & 1.18 & 0.47 \\
   ZwCl0743.5+3110 & 116.637000 &  31.022700 &  31 & 0.058 & 1.44 & 0.47 \\
   
\hline
\end{tabular}
\tablefoot{Col.~(1): System name. Col.~(2): Right ascension. Col.~(3): Declination. Col.~(4): number of spectroscopic members. Col.~(5): mean redshift. Col.~(6): $r_{200}$ in Mpc. Col.~(7): $\Delta m_{12}$.}
\end{table*}

\begin{table*}  
\caption{Global properties of the $ 0.5 < \Delta m_{12} \leq 1.0$ sample.} 
\label{tab:sample2}
\begin{center}
\begin{tabular}{lrrrrrr}
\hline 
\multicolumn{1}{l}{Name} & \multicolumn{1}{c}{R.A.} &
\multicolumn{1}{c}{Dec} & \multicolumn{1}{c}{$N_m$} & \multicolumn{1}{c}{z} &  
\multicolumn{1}{c}{$r_{200}$} & \multicolumn{1}{c}{$\Delta m_{12}$} \\  
& & & & & \multicolumn{1}{c}{[Mpc]} &  \\  
\hline
         ABELL1630 & 192.973000 &   4.579790 &  32 & 0.065 & 0.92 & 0.51 \\
         ABELL1999 & 223.511000 &  54.265700 &  36 & 0.099 & 0.94 & 0.54 \\
         ABELL0819 & 143.071000 &   9.683590 &  31 & 0.076 & 1.10 & 0.55 \\
         ABELL1783 & 205.735000 &  55.603900 &  46 & 0.069 & 0.79 & 0.59 \\
         ABELL2023 & 227.497000 &   3.003090 &  25 & 0.092 & 1.05 & 0.60 \\
         ABELL0628 & 122.535000 &  35.275300 &  54 & 0.084 & 1.37 & 0.67 \\
         ABELL1516 & 184.718000 &   5.245670 &  66 & 0.077 & 1.48 & 0.68 \\
         ABELL1809 & 208.277000 &   5.149730 &  90 & 0.079 & 1.51 & 0.72 \\
   ZwCl0027.0-0036 &   7.368420 &  -0.212620 &  35 & 0.060 & 0.96 & 0.72 \\
         ABELL1424 & 179.371000 &   5.089060 &  72 & 0.076 & 1.27 & 0.73 \\
         ABELL1728 & 200.882000 &  11.302200 &  49 & 0.090 & 1.68 & 0.73 \\
         ABELL1620 & 192.516000 &  -1.540380 &  58 & 0.085 & 1.70 & 0.74 \\
         ABELL1169 & 167.096000 &  44.150300 &  41 & 0.058 & 0.90 & 0.79 \\
 NSCJ161123+365846 & 242.808000 &  36.973400 &  44 & 0.067 & 1.00 & 0.83 \\
    RXJ1022.1+3830 & 155.656000 &  38.579200 &  62 & 0.054 & 1.23 & 0.83 \\
            WBL518 & 220.178000 &   3.465420 & 109 & 0.027 & 0.96 & 0.85 \\
         ABELL1149 & 165.740000 &   7.603880 &  28 & 0.072 & 0.73 & 0.90 \\
         ABELL2175 & 245.130000 &  29.891000 &  78 & 0.096 & 1.79 & 0.90 \\
         ABELL1346 & 175.299000 &   5.734720 &  77 & 0.098 & 1.61 & 0.91 \\
         ABELL0971 & 154.967000 &  40.988500 &  41 & 0.093 & 1.67 & 0.97 \\
         ABELL1663 & 195.719000 &  -2.517850 &  80 & 0.083 & 1.49 & 0.97 \\
         ABELL0168 &  18.740000 &   0.430810 & 105 & 0.045 & 1.21 & 0.98 \\
         ABELL2670 & 358.557000 & -10.419000 & 143 & 0.076 & 1.32 & 0.99 \\

\hline
\end{tabular}
\end{center}   
\tablefoot{Columns as in Table~\ref{tab:sample1}.}
\end{table*}    

\begin{table*}  
\caption{Global properties of the $ 1.0 < \Delta m_{12} \leq 1.5$ sample.} 
\label{tab:sample3}
\begin{center}
\begin{tabular}{lrrrrrr}
\hline 
\multicolumn{1}{l}{Name} & \multicolumn{1}{c}{R.A.} &
\multicolumn{1}{c}{Dec} & \multicolumn{1}{c}{$N_m$} & \multicolumn{1}{c}{z} &  
\multicolumn{1}{c}{$r_{200}$} & \multicolumn{1}{c}{$\Delta m_{12}$} \\  
& & & & & \multicolumn{1}{c}{[Mpc]} &  \\  
\hline
         ABELL1767 & 204.035000 &  59.206400 & 148 & 0.071 & 1.83 & 1.01 \\
         ABELL2241 & 254.933000 &  32.615300 &  38 & 0.098 & 1.64 & 1.02 \\
             FGS31 & 260.041836 &  38.834513 & 120 & 0.159 & 2.30 & 1.04 \\
         ABELL0695 & 130.305000 &  32.416600 &  20 & 0.068 & 0.94 & 1.05 \\
         ABELL0085 &  10.460300 &  -9.303130 & 271 & 0.055 & 2.03 & 1.09 \\
         ABELL2245 & 255.638000 &  33.516700 &  44 & 0.086 & 1.10 & 1.09 \\
         ABELL1750 & 202.711000 &  -1.861970 &  41 & 0.088 & 1.06 & 1.11 \\
             FGS25 & 234.961581 &  48.404745 & 117 & 0.097 & 1.67 & 1.12 \\
         ABELL0257 &  27.285000 &  13.963300 &  28 & 0.070 & 0.79 & 1.14 \\
 NSCJ152902+524945 & 232.311000 &  52.863900 &  54 & 0.074 & 1.34 & 1.16 \\
             FGS26 & 237.232728 &  44.134516 &  83 & 0.072 & 0.95 & 1.18 \\
         ABELL0724 & 134.541000 &  38.640400 &  30 & 0.094 & 0.88 & 1.20 \\
             FGS13 & 175.367899 &  10.823113 &  18 & 0.188 & 1.27 & 1.23 \\
         ABELL1564 & 188.758000 &   1.798650 &  53 & 0.079 & 1.32 & 1.25 \\
         ABELL2244 & 255.690000 &  34.061100 &  26 & 0.096 & 0.87 & 1.26 \\
   RXCJ1115.5+5426 & 168.849000 &  54.444100 &  60 & 0.070 & 1.32 & 1.28 \\
         ABELL0779 & 139.945000 &  33.749700 &  59 & 0.023 & 0.71 & 1.34 \\
             FGS19 & 203.999933 &  41.527137 &  27 & 0.177 & 1.48 & 1.35 \\
  MACSJ1440.0+3707 & 220.014000 &  37.124300 &  19 & 0.098 & 1.19 & 1.36 \\
         ABELL2428 & 334.065000 &  -9.333250 &  32 & 0.084 & 0.89 & 1.38 \\
         ABELL0152 &  17.513200 &  13.978200 &  38 & 0.060 & 1.12 & 1.40 \\
   RXCJ1351.7+4622 & 208.161000 &  46.349800 &  59 & 0.063 & 1.10 & 1.40 \\
             FGS22 & 223.495881 &  -3.524771 &  31 & 0.146 & 0.92 & 1.49 \\

\hline
\end{tabular}
\end{center}   
\tablefoot{Columns as in Table~\ref{tab:sample1}.}
\end{table*}    

\begin{table*}  
\caption{Global properties of the $\Delta m_{12} > 1.5$ sample.}
\label{tab:sample4}
\begin{center}
\begin{tabular}{lrrrrrr}
\hline 
\multicolumn{1}{l}{Name} & \multicolumn{1}{c}{R.A.} &
\multicolumn{1}{c}{Dec} & \multicolumn{1}{c}{$N_m$} & \multicolumn{1}{c}{z} &  
\multicolumn{1}{c}{$r_{200}$} & \multicolumn{1}{c}{$\Delta m_{12}$} \\  
& & & & & \multicolumn{1}{c}{Mpc} &  \\  
\hline
            RBS1385 & 215.965000 &  40.258800 &  17 & 0.082 & 0.86 & 1.60 \\
             FGS12 & 170.480324 &  40.626439 &  26 & 0.240 & 1.43 & 1.61 \\
             FGS27 & 243.629598 &  30.717775 &  76 & 0.184 & 1.37 & 1.61 \\
             FGS04 & 121.878105 &  34.011550 &  28 & 0.208 & 1.58 & 1.65 \\
         ABELL0117 &  13.966200 &  -9.985650 &  57 & 0.055 & 1.10 & 1.69 \\
   ZwCl1207.5+0542 & 182.570000 &   5.386040 &  42 & 0.077 & 1.19 & 1.69 \\
             FGS29 & 251.758645 &  26.730653 &  27 & 0.135 & 0.96 & 1.81 \\
             FGS34 & 359.562947 &  56.665593 &  35 & 0.178 & 1.07 & 1.82 \\
             FGS30 & 259.549773 &  41.188995 &  71 & 0.114 & 1.64 & 1.84 \\
         ABELL0999 & 155.849000 &  12.835000 &  25 & 0.032 & 0.57 & 1.86 \\
             FGS23 & 232.442800 &  41.755800 &  63 & 0.148 & 1.08 & 1.87 \\
         ABELL2110 & 234.962000 &  30.717800 &  25 & 0.098 & 1.27 & 1.88 \\
             FGS14 & 176.698219 &   5.974865 &  39 & 0.221 & 1.59 & 1.96 \\
             FGS17 & 191.925308 &   9.874491 &  14 & 0.155 & 0.92 & 1.96 \\
   RXCJ0953.6+0142 & 148.422000 &   1.700650 &  25 & 0.098 & 1.19 & 1.97 \\
             FGS03 & 118.184160 &  45.949280 & 102 & 0.052 & 1.03 & 2.09 \\
             FGS02 &  28.174836 &   1.007103 &  45 & 0.230 & 2.07 & 2.12 \\
             FGS20 & 212.517450 &  44.717033 &  29 & 0.094 & 0.79 & 2.17 \\
         ABELL0690 & 129.816000 &  28.844100 &  27 & 0.079 & 0.81 & 2.30 \\
             FGS28 & 249.335485 &   8.845654 &  19 & 0.032 & 0.47* & 3.28 \\

\hline
\end{tabular}
\end{center} 
\tiny *$r_{200}$ estimated using $L_X$, since FGS28 has only one member within the virial radius. We firstly estimated $r_{500}$ by using Eq. 2 from \citet{Bohringer2007}, then convert it to $r_{200}$ by using $r_{200} = 1.516 \,r_{500}$, according to \citet{Arnaud2005}. See \citet{Girardi2014} for details.
\tablefoot{Columns as in Table~\ref{tab:sample1}.}
\end{table*}    

\end{appendix}
\end{document}